%% file: main.tex
\documentclass[preprint,12pt,authoryear]{elsarticle}

\usepackage[utf8]{inputenc}
\usepackage[T1]{fontenc}
\usepackage[hidelinks]{hyperref}
\usepackage{url}
\input{anonymization.tex}
\usepackage{booktabs}
\usepackage{amsmath,amssymb}
\usepackage{siunitx}
\usepackage{graphicx}
\usepackage{placeins} 
\usepackage{doi}

\usepackage[margin=1in]{geometry}
\usepackage{microtype}
\makeatletter
\def\ps@pprintTitle{\let\@oddhead\@empty\let\@evenhead\@empty\def\@oddfoot{\hfil\thepage\hfil}\let\@evenfoot\@oddfoot}
\makeatother

\title{Funding the runners-up beats a golden ticket}

\ifanonymized
\author{Anonymous}
\hypersetup{pdfauthor={}}
\else
\author{Haining Wang}
\address{Indiana University School of Medicine, Indianapolis, IN, USA}
\ead{hw56@iu.edu}
\hypersetup{pdfauthor={Haining Wang}}
\fi

\begin{document}
\begin{frontmatter}
\begin{abstract}
A golden ticket allows one reviewer to advance a proposal that the rest of a panel would
reject. We assess its score-based rationale by comparing minority-support ranking
with other rules for allocating the same number of places. We do so on \num{10625} rejected submissions to a computer-science conference between 2017 and 2024,
using later citation percentiles as the outcome. Minority-support ranking measured the gap
between the two highest scores after adjustment for the panel mean and reviewer count.
At a budget of 10\% of the year's accepted count, panel-mean ranking selected submissions
with a mean citation percentile $7.75$ points higher than minority-support ranking
(95\% bootstrap interval $4.35$ to $11.09$). The difference was larger in the other sample periods and
citation windows examined. At that budget, minority-support ranking was equivalent to a lottery
within a margin of $\pm 5$ percentile points. Two further analyses, adjusted highest-score ranking and a matched comparison, found no clear citation advantage from minority support once the panel mean was accounted for. Ranking on score
variance also outperformed minority-support ranking in a paired comparison. Estimates
cover the \num{69.8}\% of eligible rejections whose later citations could be verified.
These comparisons identify panel-mean ranking as a benchmark for evaluating whether
golden tickets select work that would otherwise be overlooked.
\end{abstract}

\begin{keyword}
peer review \sep research funding \sep allocation rules \sep science policy \sep metascience
\end{keyword}
\end{frontmatter}

\section{Introduction}

Funding panels can reject work that later receives scholarly attention. How should funders
choose among these rejected proposals if they can support a few more? One answer is now being tried. Under a golden ticket, a
single panelist can advance a proposal the rest of the panel would reject, spending an
override that each reviewer holds once.
Villum Fonden, a Danish private foundation that supports scientific research, has given
each reviewer such a vote in its Villum Experiment program since 2017, calling it a decisive vote
\citep{bloch2026}. The technology directorate of the US National Science Foundation (NSF) described a
golden ticket in 2023 \citep{naturenews2023}, alongside a partnership to test new
merit-review mechanisms \citep{nsfifp2023}; NSF announced its new Office of Metascience
in September 2026 \citep{nsf2026}, and its fiscal year 2027 budget request proposes golden-ticket
trials \citep{nsf2027}.

The golden ticket rests on a specific claim about what happens inside a panel. Consensus
scoring is said to regress toward the unobjectionable, so that a proposal one reviewer
believes in and the others do not is exactly the proposal a mean score buries. On that
account the lone supporter holds information the aggregation step throws away, and the
override recovers it. The mechanism is not a way of funding more work; it is a claim that
a particular minority signal is informative.

The available evidence does not establish whether single-reviewer overrides improve
outcomes relative to alternative selection rules. The published study of Villum's decisive-vote program uses interviews and surveys to examine grantees' perceptions of risk \citep{bloch2026}; NSF's
proposal has no outcome evaluation attached. Studies of discretionary selection at the
Advanced Research Projects Agency-Energy (ARPA-E), where program directors choose without
a panel, have examined short-term project performance \citep{goldstein2024}. What is missing is an
evaluation of this specific rule against the alternatives an agency would otherwise spend
the same awards on. The adjacent literature is about something
else. Work on reviewer disagreement uses the variance or the range of a panel's scores
\citep{barnett2018,huang2026}, both of which are unsigned and therefore pool a panel with
one strong supporter together with a panel with one strong detractor. Work on panel
behavior establishes that panels are not simple aggregators, whether through how
applications interact within a round \citep{lawson2023} or through the implicit weights
reviewers place on criteria they score separately \citep{franzoni2026}. Work on the
predictive value of review uses the aggregate score \citep{liagha2015,fang2016}. None of
it speaks to a rule that acts on one reviewer in one direction.

Deciding whether the rule is worth adopting requires per-reviewer scores on the
submissions a panel \emph{rejected}, since those are the submissions an override acts on,
together with an observable later outcome for them. Releases of reviewer-level scores on
unfunded applications exist but are small and carry no outcomes; we use one of them below to
describe how often the configuration arises. The International
Conference on Learning Representations (ICLR) makes reviews and scores available for public submissions,
accepted and rejected alike. We use \num{21857} submissions from 2017 to 2024, of which
\num{10625} are eligible rejections.

What the paper adds is the form of the question. An agency does not ask whether
disagreement correlates with impact; it asks the citation outcomes associated with a fixed number of extra places under
one rule rather than another. For a budget
expressed as a share of the year's accepted set, we rank the rejected submissions by a
rule, take the top of that ranking, and compare their later citation outcomes. A
lottery over the same pool is the benchmark. The panel mean enters as a rule in its own
right, because extending selection down the existing score ranking provides a direct
benchmark for an override.

Three findings follow. First, panel-mean and highest-score ranking outperform the lottery at all four budgets on the two-year window. At a budget of 10\%, panel-mean ranking exceeds minority-support ranking by $7.75$ percentile points ($4.35$ to $11.09$).

Second, among submissions with observable citations, minority-support ranking is equivalent to a lottery within five percentile points at a budget of 10\% in both follow-up windows; the criterion is not met at 5\%. The adjusted highest-score ranking and matched comparison likewise show no clear advantage after accounting for the panel mean.

Third, score-variance ranking outperforms minority-support ranking in the paired comparison at a budget of 10\%, although neither rule shows a clear advantage over the lottery in the full two-year sample.

\section{Single-reviewer override and what it presumes}

\subsection{The mechanism, as agencies have built it}

Existing and proposed overrides share a structure. Villum Fonden's Villum Experiment has since 2017 given each reviewer a decisive
vote: a reviewer may
guarantee funding for one proposal in the round \citep{bloch2026}. NSF's technology
directorate described a golden ticket along the same lines in 2023 \citep{naturenews2023},
and the fiscal year 2027 budget request proposes trials across the foundation \citep{nsf2027}. In
each case the override is scarce, exercised by one reviewer, and acts against an aggregate
that would otherwise have decided the outcome on its own.

The configuration such a rule acts on is not rare in funding panels. In the one release of
per-reviewer votes on unfunded applications we could obtain, covering four sections of a
Swiss National Science Foundation call, \num{53} of \num{239} unfunded proposals, or
\num{22.2}\%, had at least one reviewer voting above the section's identified funding
boundary \citep{heyard2022}. That describes how often a golden ticket would have something to act on. It says
nothing about what acting on it returns, since the release carries no outcomes.

The scarcity matters for how the mechanism should be evaluated. An override is not a way
of funding more work in general, since an agency wanting that could move its threshold and
spend the same money. It is a claim that a particular signal, held by one reviewer and
lost in aggregation, identifies work the aggregate would wrongly reject. Evaluating it
means comparing the outcomes of selecting on that signal with those of allocating the same
number of places by another rule.

\subsection{Information beyond the panel mean}

The mechanism assumes that the position of a
single high assessment relative to its panel predicts something the panel mean does not
already predict. The premise is stronger than the observation that reviewers disagree, and
stronger than the observation that rejected work sometimes succeeds. Disagreement can be
noise \citep{pier2018}, and rejected work can succeed for reasons no panel could have seen
\citep{siler2015}, so what the premise requires is incremental information rather than any
information at all.

There are reasons to take it seriously. Panels weight criteria implicitly and unevenly
when they collapse separate scores into an overall judgment \citep{franzoni2026}, and
decisions on one application are shaped by the others in the round \citep{lawson2023}.
Evaluators mark down work at greater intellectual distance from themselves
\citep{boudreau2016}, and negative information from one reviewer carries disproportionate
weight in group evaluation \citep{lane2022}. Where risk is what divides a panel,
disagreement may reflect perceived risk: in an experiment on 605 experienced
reviewers evaluating mock review statements, higher-risk statements produced more dispersed scores \citep{gallo2022}. If novel
work also carries a fatter right tail \citep{uzzi2013,wang2017novelty}, a rule that acts
on disagreement has something to select.

The premise also has a policy motive behind it. Panels are argued to be risk-averse in a
way that funding agencies say they do not want: reviewers treat uncertainty as a defect
rather than a feature \citep{franzoni2023}, National Institutes of Health (NIH) renewal review appears to penalize
investigators who change direction \citep{azoulay2025}, and the clearest evidence that a
funding rule can raise the rate of exceptional work comes from a scheme that tolerated
failure over a long horizon rather than from any change to how proposals are scored
\citep{azoulay2011}. A golden ticket is an attempt to get that tolerance from the scoring
step itself.

Against that, a panel is also an aggregation device, and aggregation removes noise. If
individual scores are the panel mean plus error, a rule that acts on one member of the
panel is acting on the error term. Which of these two descriptions is right is an empirical
question, and simulation work that calibrates panel procedure against known proposal quality
shows the answer depends on details of the scale and the aggregation rule rather than on
general argument \citep{feliciani2022}.

\subsection{What panel scores predict, and what panels do with them}

Evidence on the predictive value of aggregate scores is mixed. At the National Heart, Lung, and Blood Institute (NHLBI), percentile ranking
across \num{6873} grants funded over thirty years discriminated between grants that went on
to produce a top-decile paper and grants that did not with an area under the curve of
\num{0.52} \citep{lauer2015}; on a narrower cohort of \num{1492} cardiovascular research project grants (R01s) there
was no monotonic association between percentile ranking and any citation measure
\citep{danthi2014}. The evidence is not uniformly null. At a biological-sciences funder, review scores
correlated moderately with later citation output across the \num{227} projects that were
funded out of \num{2063} submitted, though with wide variance and no scaling with award
size \citep{gallo2014}. A systematic
review of \num{105} studies concludes that grant peer review is at best a weak predictor of
later performance, and that its clearest documented effect is a bias against innovative
work \citep{guthrie2018}.

Where the aggregate is weak, the position of the funding line is weaker still. Bootstrapping
Australian panel scores against reviewer-level variability, \citet{graves2011} found that
\num{59}\% of the \num{620} funded grants would sometimes not have been funded had a
different set of reviewers been drawn; across all \num{2705} applications, only \num{9}\%
were funded under every draw. Reanalyzing Swiss
proposal scores with a hierarchical model, \citet{heyard2022} reach the same conclusion from
the other direction: the ranking near the line is too uncertain to justify a sharp cut-off,
which is why they propose randomizing within the band around it. An earlier study found that
the top-ranked proposal in a competition would have missed the cut-off between \num{9} and
\num{35}\% of the time depending on which pair of reviewers happened to be assigned
\citep{mayo2006}. Whether a golden ticket selects from this uncertain band is an empirical question.

Deliberation does not consistently improve agreement. Two randomly constituted panels scoring the same
\num{65} proposals agreed no better after discussion than the simple average of their
members' pre-meeting scores \citep{fogelholm2012}, and discussion in NIH-style panels raises
agreement within a panel while lowering it between panels \citep{pier2017}. What discussion
does do is move a small number of applications decisively: across a large set of panels it
shifted roughly one application in ten across a putative funding line, more often downward
than upward \citep{carpenter2015}. Those changes show that discussion can alter selection; they do not identify which
applications a reviewer would advance using an override.

One finding in this literature runs directly parallel to ours. Across \num{1561} external
reviews of \num{280} applications, the \emph{mean} reviewer score predicted the funding board's
decision far better than any individual reviewer's score, with areas under the curve of
\num{0.75} against \num{0.62}, and there was no material gain past four reviewers
\citep{sorrell2018}. That study predicts board decisions, not subsequent research outcomes. Our comparison
asks whether score-based selection predicts later citation outcomes.

\subsection{Why rejected applications are the relevant sample}

Almost all of the evidence above shares a limitation, and it is the limitation that makes
this paper possible. Scores are observed for every application; outcomes are observed only
for the ones that were funded. \citet{lindner2015} state the consequence plainly: a
correlation between scores and later citations computed among funded grants is not a valid
test of predictive validity. The funding line cuts off the bottom of the score range before
the outcome is ever observed, the awards made below the line are not a random sample of
those below it, and the funded project is often renegotiated after review. This limitation applies to the outcome studies restricted to funded applications.

The point generalizes beyond predictive validity. \citet{barnett2018} asked the question
closest to ours, whether applications on which reviewers disagreed accumulated more
citations, and found nothing for either the standard deviation or the range of scores. Their
sample was \num{227} funded applications out of \num{2063} submitted, and they note the
restriction themselves: the applications where one reviewer was enthusiastic and the others
were not are the ones most likely to have fallen below the line, and so the ones most likely
to be missing from a sample of awards. A mechanism that acts on rejections cannot be evaluated on a sample
of awards.

Agency data have also addressed the benchmark rule. When the American Recovery and Reinvestment Act briefly relaxed the
NHLBI payline, \num{165} R01 grants were funded that the ranking rule would otherwise have
rejected, alongside \num{458} funded by meeting the line. Despite shorter durations and
smaller budgets, the below-the-line grants returned a statistically indistinguishable
normalized citation impact per million dollars \citep{danthi2015arra}. This comparison of funded projects found no clear difference in citation impact per dollar;
it did not establish equivalence between projects above and below the line. It does not evaluate a single-reviewer override. It provides an empirical precedent for extending selection below the usual funding line.

\subsection{What the evidence on disagreement does and does not settle}

Disagreement does not necessarily indicate novelty. Across 49 journals, reviewer disagreement was statistically similar across levels of novelty \citep{teplitskiy2022}.

Selectors who already have discretion appear to act on disagreement. At ARPA-E, where
program directors choose rather than a panel, they preferentially fund proposals with
greater disagreement among reviewers and proposals reviewers describe as creative, and the
resulting portfolio shows no significant cost in short-term project performance
\citep{goldstein2024}. That is the closest existing analog to an override, and it does
carry outcome evidence. What it does not do is set the rule against named alternatives at a
fixed budget, which is the comparison an agency choosing between mechanisms has to make.

Closest to this paper, \citet{xie2024} study the same conference, relate reviewer
inconsistency to citations, and report a positive association including in a model
restricted to rejected submissions. That reviewer disagreement predicts citations in this
dataset is therefore not a new observation. What is open
is the decision an agency faces: whether a rule acting on a single supporter earns its
place against panel-mean ranking, a lottery, and score variance, at a budget any of
them could spend.

The lottery is in that list because funders now use it. The Health Research Council of New
Zealand allocates one scheme by ballot after a peer-review filter, and applicants to it
favor randomization by roughly two to one \citep{liu2020}; surveyed scientists more
generally reject pure randomization but accept it once peer review has screened the field
\citep{philipps2022}. Proposals to combine review with a draw are now specific enough to
have a design taxonomy \citep{shaw2023,fang2016lottery} and a formal treatment of where in
the ranking the draw belongs \citep{heyard2022}, and the case that randomization protects
unconventional work has been made in detail \citep{avin2019}. These practices motivate the lottery benchmark. We draw from all eligible rejections;
funder lotteries generally draw from applications that have passed a quality threshold.

\subsection{Research questions}

\begin{enumerate}
\item With the same number of extra places, do the panel mean and the highest score
  select rejected submissions with better citation outcomes than a lottery or minority-support ranking?
\item Does one reviewer's unusually strong support predict more citations after
  accounting for the panel mean?
\item Does selecting for disagreement among reviewers perform better than selecting
  for one reviewer's unusually strong support, the signal a golden ticket is intended to capture?
\end{enumerate}

\section{Setting and data}

\subsection{Why conference review}

ICLR provides the reviewer scores and rejection decisions needed for this comparison. Submissions are
scored individually by three to five reviewers on a numeric scale, and these reviews inform acceptance decisions. An experiment at the Conference on Neural Information Processing Systems documents substantial variability in conference-review scores \citep{cortes2021}. Most importantly for this
question, rejection is observable at the level of the individual score, which permits a comparison of rules within the rejected pool when later outcomes can be linked.

The dataset is already used this way. Studies of it have examined how borderline
submissions are treated \citep{ibrahim2026}, whether authors' own rankings of their
submissions predict impact \citep{su2025}, and what distinguishes the rejections that
later matter \citep{huang2026}.

\subsection{Dataset}

The Berens Lab ICLR dataset provides public submission records, decisions and per-reviewer scores \citep{berensdataset}, making
comparisons of selection rules possible and providing a widely used dataset for studying review \citep{wang2022openreview}. We build on a
public linkage of those submissions to bibliographic records \citep{snor2025}, extended
with additional bibliographic matching. We take \num{21857} submissions, of which \num{10625} are explicit
rejections carrying at least two numeric scores. Supplementary Section~\ref{sec:s-selection} and Figure~\ref{fig:s-flow} trace sample selection; Supplementary Section~\ref{sec:s-characteristics} and Table~\ref{tab:s-coverage} describe the eligible pool. Supplementary Section~\ref{sec:s-dataset} details dataset preparation. Withdrawn and desk-rejected submissions
are held out of the main analysis and treated separately, because a withdrawal after
reviews is a different event from a rejection.

The main analysis excludes the 2020 edition, which used four score values
(1, 3, 6 and 8). We include that edition in a sensitivity analysis of the ranking rules. The edition contributes no matched pairs (Supplementary Section~\ref{sec:s-2020}).

\subsection{Score normalization and the estimated acceptance threshold}

Score scales and their verbal anchors change across editions, and the change is not
cosmetic: the anchor attached to a given number moved at the 2022 review-form revision, so
a fixed numeric threshold is stricter in early years and looser in late ones. We therefore
anchor on the bar each year's submissions actually faced. Within each year we fit
acceptance on the panel mean and take the mean-score value at which the fitted probability
of acceptance is one half. This estimated acceptance bar defines the bar-clearing indicator below.
The logistic fit includes all submissions with at least two scores and treats decisions other than acceptance as non-acceptance. Its probability midpoint is an operational threshold, not an observed funding line. The ranking rules are evaluated within each year.

Supplementary Table~\ref{tab:s-scores} reports panel means by edition.
Individual scores are standardized using the mean and standard deviation of the scores within each year. Reviewers cannot be linked across
submissions, because the platform issues a fresh anonymous alias for every assignment, so
no reviewer-level leniency adjustment is possible and none is attempted.

\section{Methods}
\label{sec:methods}

\subsection{Allocation rules}

Each rule orders the rejected pool within a year; the top of that ordering is what the rule
would have advanced. We examine minority support through the gap between the two highest scores, the highest score after adjustment, and a matched comparison based on whether any score reaches the estimated acceptance bar. The main ranking comparison uses four rules and a lottery benchmark:

\begin{description}
\item[Minority-support ranking.] The gap between the highest and second-highest standardized score,
  after removing its linear association with the panel mean and the number of reviewers. This rule favors a submission with one unusually enthusiastic reviewer. It measures the support that a golden ticket is intended to act on.
\item[Panel mean.] Rank rejected submissions by their mean standardized score. This extends selection
  down the existing score ranking.
\item[Highest score.] The maximum standardized score in the panel: submissions with the most favorable single review rank first. In a supplementary analysis, we also rank the part of this score not explained by a linear fit to the panel mean and reviewer count. This tests whether the highest score adds predictive information beyond those two variables (Supplementary Section~\ref{sec:s-adjusted}, Figure~\ref{fig:s-adjusted}, and Table~\ref{tab:s-adjusted}).
\item[Score variance.] The variance of the panel's standardized scores. It ranks submissions with the greatest disagreement first, whether that disagreement comes from an unusually high or unusually low score.
\item[Lottery.] A uniform draw of the same size from the same pool. We calculate its expected
  mean directly within each bootstrap resample and use it as the benchmark.
\end{description}

A separate construction, the bar-clearing indicator, is used for matching rather than ranking.
A rejected submission clears the bar when at least one reviewer scored it at or above the
year's estimated acceptance threshold. This indicator captures one feature of the mechanism: somebody on
the panel thought the submission was fundable. It is an indicator, not an ordering, so it is
evaluated by matching rather than by a citation gain at a fixed budget, and it therefore does not enter the
budgeted comparison against panel-mean ranking.

Each submission that clears the bar is matched one-to-one to a rejection from the same
edition with the same number of reviewers and the same panel mean, without replacement.
This compares citation outcomes at the same average rating. Intervals are bootstrapped
over the fixed matched pairs. Supplementary Section~\ref{sec:s-matching} reports a sensitivity analysis
that allows a small difference in panel means to retain more pairs.

\subsection{Outcome and citation window}

The outcome is the citation percentile: a submission's position in the citation distribution
of submissions from the same year, with higher percentiles indicating more citations.
Citations are counted over a fixed follow-up window from the conference year. The main window is two
years, which retains every included edition; a three-year window is reported alongside and restricts
the sample to editions through 2022. We count citations using the June 26, 2026 version of OpenAlex, an open database of
scholarly publications \citep{priem2022openalex}. A paper citing more than one version of
a submission counts only once. Supplementary Section~\ref{sec:s-dataset} describes data preparation and linkage;
Supplementary Section~\ref{sec:s-window} checks citation dates around the decision.
Percentiles are computed within year, so a cohort's overall citation level does not enter
any comparison.

\subsection{What we measure, and how the intervals are built}

Let $B$ denote the budget, expressed as a percentage of each year's accepted count,
with $B \in \{1,2,5,10\}$, rounding each year's number of places up to the next integer. Let $r$ identify the selection rule. The set $S_r(B)$
contains the rejected submissions selected by rule $r$ at that budget. For any selected
set, $\overline{y}$ denotes its mean citation percentile. Let $\mu_L(B)$ denote the
expected mean citation percentile under a uniform lottery selecting the same number
of rejections from each year. We calculate this expectation directly from the eligible
pool, rather than from a single random draw. Only available citation outcomes enter these means; Supplementary Section~\ref{sec:s-weights} describes the resulting year weights. Supplementary Section~\ref{sec:s-implementation} describes tie handling. The citation gain, $\Delta_r(B)$, is the
difference between the rule's mean and this lottery expectation, in percentile points:
\[
\Delta_r(B) \;=\; \overline{y}\bigl(S_r(B)\bigr) \;-\; \mu_L(B).
\]
Intervals are percentile
bootstrap over submissions within year, \num{2000} resamples. Adjusted highest-score ranking is
refit inside each resample, so the uncertainty in that fit is carried. The minority-support
ranking uses an adjustment computed once on the full sample and held fixed across
resamples; its intervals therefore do not carry the uncertainty in that step.

Comparisons between two rules use the same bootstrap draws for both terms, so the interval
on a difference reflects the correlation between rules evaluated on one pool. This is what
separate intervals cannot deliver: two rules can differ from each other detectably while
neither differs detectably from the lottery.

\subsection{Equivalence margin and interpretation}

We use a fixed equivalence margin of $\pm 5$ citation percentile points for all rule
comparisons. Relative to the standard deviation of citation percentiles among eligible
rejections with observed outcomes, this is approximately $\pm 0.19$ standard deviations,
close to the conventional small-effect reference of $0.2$ \citep{williams2014,lakens2018}.
This comparison puts the margin on a familiar scale; it does not establish how much a
funding agency should value a difference of this size. Supplementary Section~\ref{sec:s-margin} and Table~\ref{tab:s-margin} define
the reference standard deviation and report sensitivity to alternative margins.

We evaluate equivalence where the 95\% interval half-width is below five points, which holds at budgets of 5\% and 10\% (Supplementary Section~\ref{sec:s-margin}). Equivalence requires the entire 90\% bootstrap interval to fall within the fixed margin
\citep{lakens2017}. These comparisons concern submissions with observable citation outcomes. The margin was fixed after preliminary results had been examined.

\subsection{Outcome resolution and missingness}

Not every rejected submission has a findable citation record: \num{69.8}\% of the eligible
rejections do, rising to \num{85.9}\% for editions through 2022. Every estimate below covers
those submissions and not the rest. We report three assumptions about the remainder, assigning unresolved
submissions the lowest percentile, the resolved pool mean, or the percentile corresponding
to zero citations, and recomputing the benchmark under each. These are assumptions, not
bounds.

We separately report worst-case bounds on what the missing outcomes could do, a calculation
of how far the unresolved selections would have to move to change the conclusion, and an
audit of a stratified sample of unmatched rejections. An unmatched submission is not a submission that
was never published, and nothing here treats it as one.

\section{Results}
\label{sec:results}

\subsection{Panel-mean ranking outperforms the lottery and minority-support ranking}

\begin{figure}[!htbp]
\centering
\includegraphics[width=\textwidth]{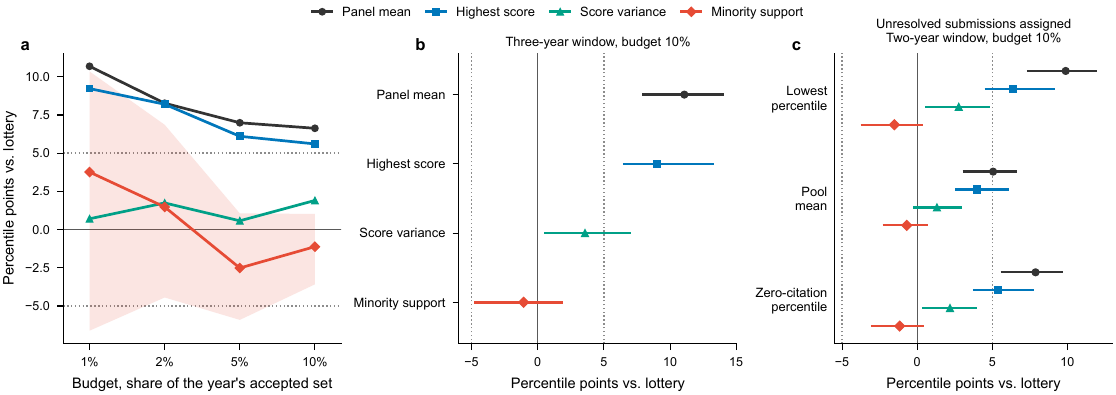}
\caption{\textbf{Citation gains under each rule.} Mean citation percentile of the selected set minus
that of a lottery over the same pool, in percentile points. Dotted reference lines mark the
$\pm 5$ percentile-point margin. The shared legend identifies the rules in all panels. \textbf{a}, citation gain against budget
on the two-year window; the colored band is the 95\% interval for minority support. At budgets of 1\% and 2\%, the half-width of that interval exceeds five points, so equivalence is not evaluated there. \textbf{b}, the three-year window at
a budget of 10\%, with 95\% intervals. \textbf{c}, the two-year window at a budget of
10\% under three assumptions about submissions with no citation record, with 95\%
bootstrap intervals conditional on each assumption. These intervals do not account for
uncertainty about which missingness assumption is correct. Panels \textbf{a} and \textbf{b} are the same comparison on
different follow-up windows and sets of editions rather than independent evidence, and panel \textbf{c} is a
sensitivity analysis on the rightmost budget of \textbf{a}.}
\label{fig:results}
\end{figure}

Panel-mean and highest-score ranking both outperform the lottery at all four budgets on the two-year window, with smaller gains as the budget increases. Panel-mean ranking yields gains of
$10.69$, $8.27$, $6.99$ and $6.63$ percentile points as the budget widens from 1 to 10\%
of the year's accepted set, and the interval excludes zero at all four
(Figure~\ref{fig:results}a; Supplementary Table~\ref{tab:s-yields}). The single highest score, ranked as it
stands, yields gains of $9.23$, $8.20$, $6.11$ and $5.61$, and its interval excludes zero at all four
as well. At the 10\% budget, the 95\% intervals are $4.11$ to $8.66$ points for panel-mean ranking and $3.59$ to $8.47$ for highest-score ranking. On the three-year window the two rules yield gains of $11.08$ ($7.90$ to $14.10$) and $9.02$
($6.44$ to $13.29$) at a budget of 10\%.

\FloatBarrier

Panel-mean ranking therefore serves as the benchmark for minority-support ranking. At a budget of 10\%, panel-mean ranking selects submissions with a mean citation percentile
$7.75$ points higher than minority-support ranking (95\% bootstrap interval $4.35$ to
$11.09$). Highest-score ranking exceeds minority-support ranking by $6.72$ points ($4.34$ to $10.24$). Supplementary Table~\ref{tab:contrast} gives the paired comparisons across rules and samples (Supplementary Section~\ref{sec:s-contrasts}).

\subsection{Minority-support ranking meets the equivalence criterion}

We examined minority support using two ranking rules and a separate matched comparison.
None showed clear evidence of an advantage after accounting for the panel mean.

Minority-support ranking yields a difference of $-1.12$ percentile points at a budget of 10\%, with a 95\%
interval from $-3.58$ to $1.02$, and $-2.51$ ($-5.91$ to $1.06$) at 5\%. At a budget of
10\% the intervals meet the equivalence criterion. The 90\% interval lies entirely inside the $\pm 5$ margin on the
two-year window ($-3.30$ to $0.63$) and again on the three-year window ($-4.25$ to $1.36$),
which is the interval form of declaring the two rules practically equivalent.

This conclusion depends on the budget, equivalence margin and available outcomes. It holds at a
budget of 10\% and not at 5\%, where the 90\% interval crosses the margin ($-5.47$ to
$0.54$).
The comparison covers the submissions whose outcomes
we can see. The two-year window runs through
2024 and the three-year window stops at 2022; these overlapping samples are not independent replications. The equivalence criterion concerns average citation outcomes, not whether the rules
select the same submissions. No comparison at any budget or window establishes inferiority either.

After adjustment for the panel mean and reviewer count, highest-score ranking showed no
clear gain over the lottery at a budget of 10\% ($-1.29$ percentile points, 95\% bootstrap interval
$-3.42$ to $1.23$; Supplementary Figure~\ref{fig:s-adjusted}). This adjustment removes only linear associations.

The matched comparison asks whether a reviewer scoring a submission above the estimated
acceptance threshold predicts more citations at the same panel mean. Of eligible rejections, \num{6582} (61.9\%) clear this bar (Supplementary Table~\ref{tab:s-coverage}). Matching within year
and reviewer count yields \num{1012} pairs, of which \num{698} have citation outcomes
available for both submissions. In these pairs, the difference is $+0.17$ percentile
points ($-2.60$ to $2.76$), providing no clear evidence of a citation advantage.
The comparison applies to the submissions that could be matched; Supplementary Section~\ref{sec:s-matching}
examines the effect of allowing a small difference in panel means.

\subsection{Score variance outperforms minority-support ranking}

Reading the panel's disagreement without regard to direction also beats reading it in one
direction only. Ranking on score variance, which rises whether the odd reviewer out is
enthusiastic or hostile, outperforms minority support by $3.02$ percentile points ($0.51$ to $5.98$) at a
budget of 10\%, and by $4.47$ ($1.76$ to $7.94$), $4.21$ ($1.18$ to $8.37$) and $4.64$
($1.46$ to $8.88$) in the other three samples listed in Supplementary Table~\ref{tab:contrast} (Figure~\ref{fig:mechanism}a). At 5\% the
difference is $3.09$ ($-0.26$ to $7.28$), the same size with an interval that just includes
zero.

\begin{figure}[!htbp]
\centering
\includegraphics[width=\textwidth]{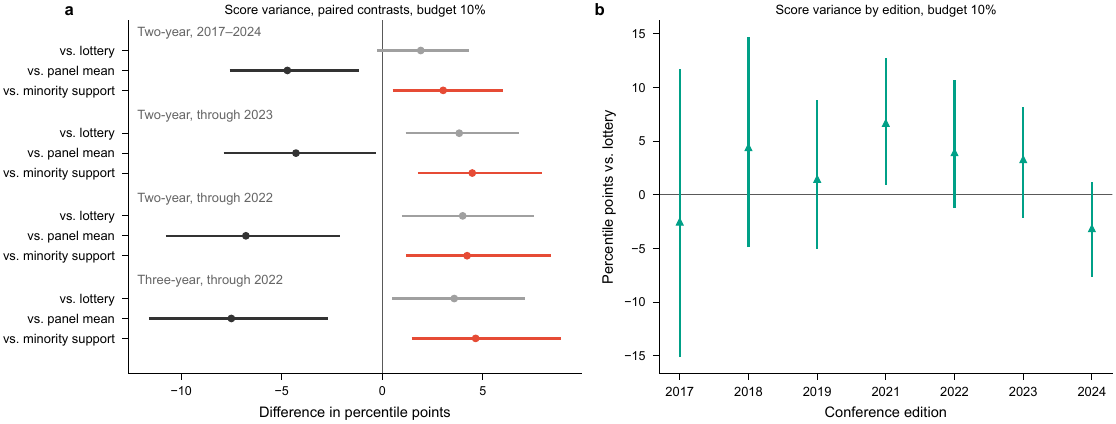}
\caption{\textbf{Score-variance ranking across comparisons and editions.}
\textbf{a}, paired contrasts at a budget of 10\% against the lottery, panel-mean ranking
and minority-support ranking in four nested samples. \textbf{b}, gains against the lottery
within each edition at a budget of 10\%. Points are estimates and whiskers are 95\%
bootstrap intervals. In panel \textbf{a}, color identifies the comparison rule: gray for lottery, black for panel mean, and red for minority support, consistent with Figure~\ref{fig:results}. In panel \textbf{b}, green denotes score-variance ranking.}
\label{fig:mechanism}
\end{figure}
\FloatBarrier

Two things these comparisons do not show. They do not show that score variance carries
information independent of the panel mean: the same contrast puts variance $4.72$ points
($1.18$ to $7.59$) below the panel-mean rule, so whatever it reads, the aggregate reads
more of it. And they do not show that either disagreement rule beats a lottery. Variance
against the lottery is $1.91$ ($-0.26$ to $4.29$) and minority support is $-1.12$ ($-3.58$
to $1.02$); the two rules separate from each other while neither separates from chance.
The paired comparison accounts for the correlation between the two rule estimates.

The variance rule's own record is uneven and we report it as exploratory. It clears zero
against the lottery on the three-year window, $3.58$ ($0.48$ to $7.07$), and on editions
through 2023, $3.82$ ($1.15$ to $6.77$), and does not on the full sample. Adding the 2024
edition alone moves it across: that edition's single-year estimate is $-3.12$ ($-7.63$ to
$1.21$), and it is the edition with the least mature citation window: it closes on
December 31, 2025, six months before the bibliographic snapshot, whereas early editions had four or more years
after their citation windows closed for database records to accumulate. All two-year
outcomes nevertheless count citations within windows of the same length. The per-edition series is otherwise not a decline. It runs
$-2.50$, $4.43$, $1.48$, $6.69$, $3.96$, $3.31$ and $-3.12$ from 2017 to 2024, with only
2021 ($0.93$ to $12.78$) clearly positive and the rest wide. These comparisons do not identify the cause of the difference between editions.
One measured feature also changes over time: the median number of
reviewers per rejection rises from 3 in 2017--2019 to 4 from 2021.

Variance and the panel mean are not selecting the same submissions. At a budget of 10\% on
the main window they share \num{67} of \num{686} selections, under a tenth, which describes how differently the two rules select. What the overlap cannot do is settle whether
variance carries anything of its own; a low share would not prove independent value and a
high share would not disprove it, since a few substituted submissions can carry the whole
difference.

\subsection{Sensitivity analyses}

Including the 2020 edition preserves the advantage of panel-mean ranking over minority
support at the 10\% budget in both citation windows
(Supplementary Section~\ref{sec:s-2020} and Table~\ref{tab:s-2020}).

The three-year window retains the advantage of panel-mean and highest-score ranking.
Including withdrawn submissions with scores gives minority support a gain of $0.98$
percentile points ($-1.55$ to $3.14$), compared with $5.51$ ($3.25$ to $7.84$) for panel-mean
ranking and $5.99$ ($3.54$ to $8.49$) for highest-score ranking. The latter two exchange
places, but both retain higher estimated gains than minority support
(Supplementary Section~\ref{sec:s-missing} and Table~\ref{tab:robust}).

Under the three missing-outcome assumptions, minority-support gains range from $-1.52$
to $-0.69$ points, while panel-mean gains range from $5.05$ to $9.88$
(Figure~\ref{fig:results}c). These scenarios preserve the observed pattern. Without
assumptions about the missing outcomes, however, the bounds are too wide to establish
equivalence for the full rejected pool. Supplementary Section~\ref{sec:s-missing} reports the bounds
and the changes in missing outcomes needed to alter the comparison.

\section{Discussion}
\label{sec:discussion}

An agency considering a golden ticket is choosing between rules for spending a fixed number
of extra awards. Among rejected submissions, panel-mean ranking selects work cited
above the lottery benchmark at every budget on the two-year window and at a budget of 10\% on the three-year window. Ranking
on the single highest score does the same. The estimated gain is largest where the budget is tightest.

Minority-support ranking and adjusted highest-score ranking showed no clear advantage
over the lottery. The separate matched comparison likewise did not establish an advantage
for the bar-clearing indicator. These results concern score-based measures of minority
support rather than reviewers' decisions to use a golden ticket. The ranking analyses remove linear associations with the panel mean; exact matching compares submissions with identical panel means but retains fewer submissions. These results do not distinguish an absent increment from one too small to detect in these samples.

At a budget of 10\% of the accepted count, panel-mean ranking selected submissions with
mean citation percentiles between $7.75$ and $12.15$ points above minority-support ranking,
depending on the follow-up window and the editions included. Each paired interval excluded
zero. This comparison supports using panel-mean ranking as a benchmark in evaluations of
golden-ticket selection.

The paired comparison also distinguishes the two disagreement rules examined here.
Score-variance ranking outperforms minority-support ranking at the largest budget. This is a
difference in performance between two rules, not evidence that score variance carries
information the panel mean does not, and the same comparison places variance well below the
panel mean. Neither disagreement rule shows a clear gain over the lottery in the full sample.
The difference between them is evidence about these two score-based rules, not a direct
comparison of funding mechanisms. In these data, reading disagreement in both directions performs better than selecting for unusually strong support alone. The comparison does not support the proposed advantage of that one-sided signal.

\subsection{Implications for evaluating golden tickets}

Panel-mean ranking uses existing scores, whereas a golden ticket asks reviewers to make an
additional decision about when to use their discretion. We did not measure the costs or
benefits of administering either procedure. Our comparisons hold the number of places
constant; applying this comparison to funding also requires accounting for differences in
award size. The results identify a simple selection benchmark, not a cost-effectiveness
estimate.

The ranking and matched comparisons leave open whether reviewers use information absent
from their scores when exercising a real golden ticket. Evaluating actual ticket use requires agency data. Villum Fonden has operated a decisive vote since 2017 \citep{bloch2026},
and could provide a setting for such an evaluation. The decisive vote is exercised rather than simulated, so the comparison is between
proposals actually advanced by one reviewer and proposals that were not, which removes the
step where we impute what a rule would have done. Funding is observed rather than imputed, so the
quantity on offer is about awards actually made rather than about the predictive content of
a score. Identifying the effect of the award would still need a design: the proposals one
reviewer advanced are not comparable to those nobody advanced without one. And a funder sees its
rejected applicants and can follow them, as studies with agency data have done
\citep{mancuso2026,lawson2023}. An agency evaluation could compare ticket use with explicit alternatives at the same
budget and specify meaningful differences before examining the outcomes.

\subsection{Limitations}

Roughly three in ten eligible rejections have no findable citation record, and every
citation estimate rests on the ones that do. The three scenarios we report are assumptions
about the unresolved submissions rather than bounds on them, and bounds that assume nothing
are uninformative. An audit of the unmatched pool returned uncertainty as its dominant
verdict rather than absence. A reader who believes the unresolved submissions are exactly
the ones a lone supporter was right about is not answered by anything here.

Reviewers cannot be linked across submissions, because the platform issues a fresh alias
for every assignment. We therefore cannot adjust for how lenient a given reviewer is, and
we cannot reproduce the scarcity that makes a golden ticket a real constraint, which is
that each reviewer gets one. Our budget binds at the level of awards rather than reviewers.

Precision also depends on the number of places allocated. At the smallest tested budget, 1\%, the intervals for panel-mean and highest-score ranking against minority-support ranking include zero; at 2\%, 5\% and 10\%, they exclude zero (Supplementary Table~\ref{tab:s-budget-contrasts}).

Scores are those publicly available when the source records were collected and may include revisions during discussion. Without revision histories, we cannot distinguish initial enthusiasm from support expressed after discussion.

The outcome is citation percentile within a submission-year cohort. It is the outcome the
dataset supports and it is not the outcome the mechanism was designed to produce. A golden
ticket is argued for on the strength of the rare results agencies call transformative, and
our analysis of mean citation percentiles does not establish performance in the extreme tail. A null on the mean is not a null on the tail.

The paired comparisons, the per-edition series and the variance rule's record are
exploratory. Each interval describes one comparison; we do not adjust for testing several
comparisons at once.

Finally, this study compares conference-review scores rather than funding decisions; it
does not observe actual golden-ticket use or estimate the effect of receiving funding.

\section{Conclusion}

A golden ticket lets one reviewer advance work the panel would not fund. Allocating the same
number of places using panel-mean ranking selects submissions cited $7.75$ percentile points higher than those minority-support ranking selects, and the gap is wider in every other sample period and window we
can form. Minority-support ranking meets the equivalence criterion against a lottery at a budget of 10\%, within a margin of five percentile points and among rejections whose later citations can be verified; the criterion is not met at 5\%. Three analyses did not establish an advantage from minority support after accounting
for the panel mean, and score-variance ranking outperforms minority-support ranking at this budget. These findings support panel-mean ranking as a benchmark for prospective evaluations
of golden-ticket selection.

\section*{Data and code availability}
The conference submissions, reviews and per-reviewer scores analyzed here are public on
OpenReview. Analysis code, aggregate results, and the versioned
\emph{\studydataset} are maintained together \studyrepository. The dataset includes submission metadata,
review scores, decisions, bibliographic identifiers, version relations, annual citation
counts, and structured matching evidence. It includes the 2020 edition and recovered
records outside the sample used in these analyses. Original code and dataset contributions use
the MIT License; upstream fields retain their source licenses and attribution, as
specified in the repository.

\section*{Ethics}
The study uses public records of a scientific review process and no human participants.
Reviewers are anonymous in the source data and cannot be linked across submissions, and no
attempt was made to identify them.

\ifanonymized\else
\section*{Acknowledgments}
The author thanks Dr. Jing Su of Indiana University School of Medicine for insights
shared during discussions of recent funding practices that inspired the idea for this paper.

\fi

\section*{Competing interests}
The author declares no competing interests.

\ifanonymized\else
\section*{CRediT authorship contribution statement}
\textbf{Haining Wang:} conceptualization, methodology, software, formal analysis,
investigation, data curation, writing: original draft, writing: review and editing,
visualization.
\fi

\bibliographystyle{elsarticle-harv}
\bibliography{references}

\clearpage
\input{supplement/content.tex}
\end{document}

%% file: anonymization.tex
\newif\ifanonymized
\anonymizedfalse

\newcommand{\studyrepository}{%
  \ifanonymized
    in the study repository (link withheld for anonymous review)%
  \else
    at \url{https://github.com/Wang-Haining/wonka}%
  \fi}
\newcommand{\studydataset}{%
  \ifanonymized study dataset\else iclr-golden-ticket, 2017--2024 dataset\fi}

%% file: supplement/content.tex
\setcounter{section}{0}
\setcounter{subsection}{0}
\setcounter{figure}{0}
\setcounter{table}{0}
\renewcommand{\thesection}{S\arabic{section}}
\renewcommand{\thefigure}{S\arabic{figure}}
\renewcommand{\thetable}{S\arabic{table}}
\renewcommand{\theHsection}{supp.\arabic{section}}
\renewcommand{\theHfigure}{supp.\arabic{figure}}
\renewcommand{\theHtable}{supp.\arabic{table}}
\section*{Supplementary Appendix}
\section{Sample selection}
\label{sec:s-selection}
\begin{figure}[ht]\centering\includegraphics[width=\textwidth]{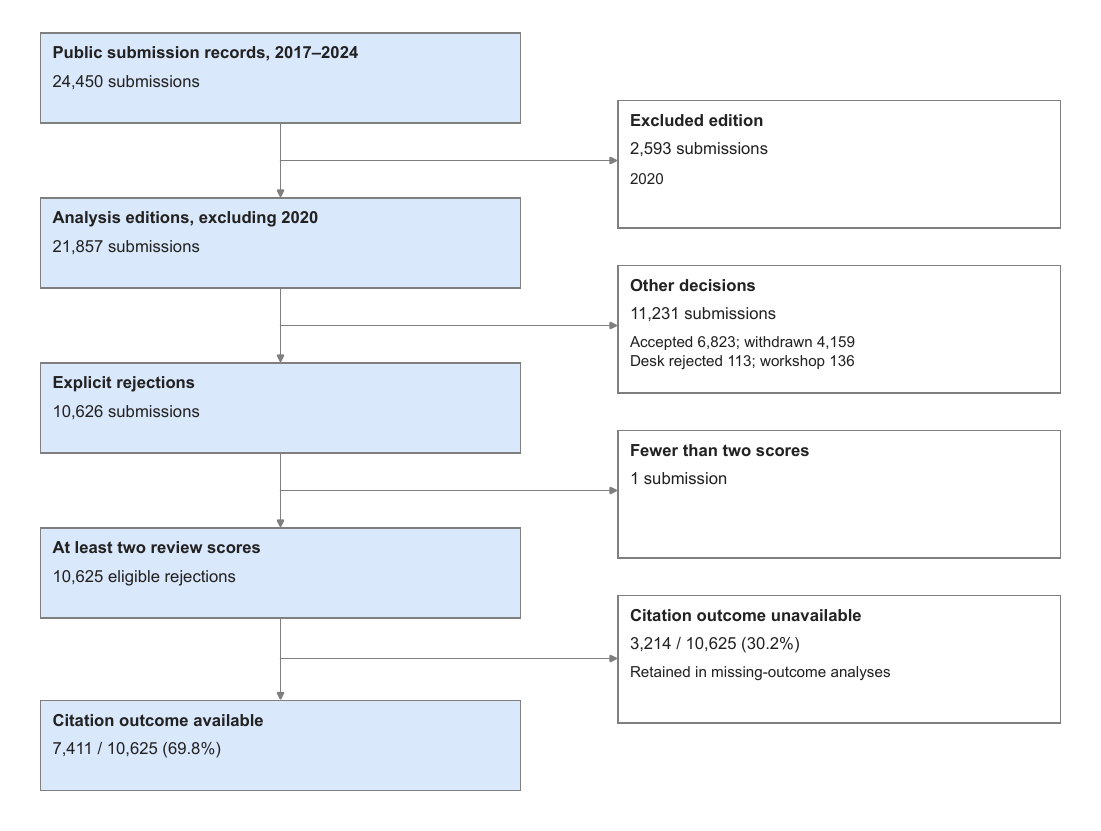}
\caption{Selection of the main two-year analysis sample. Both outcome percentages use the 10,625 eligible rejections as their denominator: 7,411 (69.8\%) have a two-year citation outcome and 3,214 (30.2\%) do not. Unavailable citation outcomes are retained as missing and included in sensitivity analyses; they are not coded as zero.}
\label{fig:s-flow}
\end{figure}

The analysis uses 24,450 source records, fixed before recovered records were added to the released dataset. The separately released dataset also
contains 26 recovered records; the recovered records do not enter these analyses, and the 2020 edition enters only
the sensitivity analysis in Section~\ref{sec:s-2020}. Publicly retrievable records are not an independently verified census of every
submission. Figure~\ref{fig:s-flow} shows sample selection.

\clearpage

\section{Sample characteristics}
\label{sec:s-characteristics}
\begin{table}[ht]\centering\small
\caption{Citation-outcome coverage and characteristics of eligible rejections. Panel A reports counts and row percentages: each available or unavailable count is divided by the eligible count in the same row. Panel B reports standard deviations (SD) and describes each column separately; the bar-clearing percentages use column totals of 10,625, 7,411 and 3,214. An unavailable outcome means that the two-year citation outcome cannot be calculated from the linked records; it does not mean zero citations or no publication.}
\label{tab:s-coverage}
\input{supplement/tableS1.tex}
\end{table}
\begin{table}[ht]\centering\small
\caption{Panel-mean score within each edition. SD denotes standard deviation. Raw scales differ across editions, so no pooled mean score is presented.}
\label{tab:s-scores}
\input{supplement/tableS2.tex}
\end{table}

\clearpage

\section{Dataset preparation and bibliographic linkage}
\label{sec:s-dataset}
The submission records, review scores and decisions come from the Berens Lab ICLR
dataset \citep[version iclr25v2;][]{berensdataset}, assembled from public OpenReview records. Existing bibliographic links come from SNOR \citep{snor2025}. We retain one row per submission
identifier and year. The analysis uses the original source records; the public release
also includes the recovered records described in Section~\ref{sec:s-selection}. Repeated submissions of
the same manuscript remain separate submissions, linked through verified bibliographic
identifiers. They are not counted as distinct manuscripts in the release's version relations.

Scores are those retained in the Berens Lab snapshot. The upstream scripts read the public rating or recommendation fields available when records were collected and carry earlier editions forward into later releases. They do not retain score-revision histories, so these data do not distinguish initial scores from scores revised during discussion. Scores are parsed within each year's review form. The analysis separates accepted,
explicitly rejected, withdrawn and desk-rejected submissions, and includes in the rejected
pool only submissions with at least two numeric scores. Sections~\ref{sec:s-selection} and~\ref{sec:s-characteristics} report the selection
counts and citation availability; the main text explains the exclusion of the 2020 edition.

We extend existing bibliographic links using persistent identifiers and title and author
evidence from OpenAlex, Semantic Scholar, arXiv, Crossref and DataCite. Candidate links
are checked for identity and document type. Changed titles and conflicting versions
require additional source evidence. Related papers, datasets, software and other nonpaper
outputs are not treated as versions of the submitted manuscript. The release retains
matching evidence, unresolved cases and the history of corrections.

All citation counts use the same OpenAlex database version, dated June 26, 2026
\citep{priem2022openalex}. For each submission, we identify papers citing any of its
verified versions and count each citing OpenAlex work once. We then count citations by
calendar year to construct the two-year and three-year outcomes. We do not add citation
totals across databases or across versions. An unavailable citation outcome stays missing;
a zero is recorded only when the citation data support a count of zero in the specified window.

The \emph{\studydataset} contains five linked data tables.
The submission table has 24,476 records and 42 fields, covering conference year,
submission identifier, title, publicly recorded authors, decision, individual review
scores, bibliographic identifiers, linkage status, citation availability, and source
provenance. It includes the 24,450 original source records and 26 recovered records.
The combination of conference year and submission identifier is the unique key;
verified work-family identifiers connect repeated submissions without merging them.

The accompanying tables describe external versions (13 fields), structured matching
evidence (22 fields), annual citation counts (11 fields), and links between cited and
citing works (3 fields). A separate 10-field analysis table retains the 24,450 original
records, the score measures used in the analyses, and citation outcomes used for reproduction.
Bibliographic identities are resolved for 19,864 submissions (81.2\%), and citation
records are available for 18,657 (76.2\%); both percentages use all 24,476 submissions
as the denominator. The repository provides field definitions, source attribution,
reuse terms, and checksums for each distributed table.

\subsection{Quality checks of bibliographic links} We drew an edition-stratified sample of 150 from the
2,842 rejections with no verified bibliographic identity (a subset of the 3,214 without a citation outcome) and adjudicated each under a fixed
protocol. Eight yielded a preprint or published version of the same work, a weighted rate
of 3.1\% (0.3 to 19.6); 107 produced a candidate whose identity
remained uncertain; 35 completed searches found no candidate, a weighted share of 31.2\%, and none was blocked by an access barrier. The dominant verdict is
uncertainty, not absence: these are not submissions shown never to have been published, and
we do not treat them as such. A separate sample of 30 newly accepted links found no
false positives, which bounds rather than certifies the matching rule. The eight recovered
identities are reported separately and do not enter the comparisons in the main text.

\FloatBarrier

\section{Highest score after adjustment}
\label{sec:s-adjusted}
\begin{figure}[ht]\centering
\includegraphics[width=.72\textwidth]{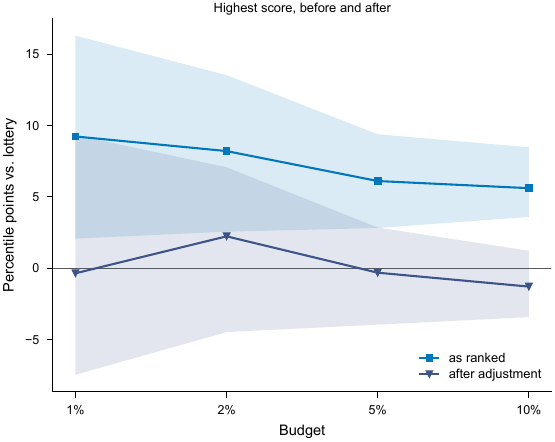}
\caption{Highest-score ranking before and after linear adjustment for panel mean and reviewer count. Points show citation gains against the lottery; bands are 95\% bootstrap intervals. Two-year window, 2017--2024 excluding 2020. No equivalence margin is drawn because this panel assesses the adjustment rather than equivalence.}
\label{fig:s-adjusted}
\end{figure}

Within each edition, we fit ordinary least squares among all eligible rejected submissions,
with the highest score as the dependent variable and an intercept, panel mean and reviewer
count as predictors. We rank the residuals and apply the same budgets as in the main analysis.
The fit is repeated within each bootstrap resample. This differs from minority-support
ranking, whose adjustment coefficients are held fixed across resamples.
After linear adjustment for the panel mean and reviewer count, highest-score ranking
yields a difference of $-1.29$ ($-3.42$ to $1.23$) at a budget of 10\%, and every adjusted highest-score estimate at
every budget has an interval covering zero (Figure~\ref{fig:s-adjusted}). The raw highest-score advantage was not evident after this adjustment. The adjustment is
linear: the panel mean explains $0.589$
of the within-year variance of the maximum pooled across editions, and $0.600$ once the
number of reviewers is added, with the fit falling from $0.766$ in 2017 to $0.556$ in 2024.
The residual is the part of the maximum those two variables do not explain linearly.
The analysis found no clear advantage in this remaining part of the score; it does not
establish independence from the panel's judgment.

\begin{table}[ht]\centering\small
\caption{Highest-score ranking after adjustment: citation gain against the lottery (95\% interval).}
\label{tab:s-adjusted}
\setlength{\tabcolsep}{4pt}
\begin{tabular}{lrrrr}\toprule
Window & 1\% & 2\% & 5\% & 10\% \\
\midrule
Two-year & $-0.35\; (-7.46, 9.33)$ & $2.23\; (-4.47, 7.07)$ & $-0.31\; (-3.95, 2.85)$ & $-1.29\; (-3.42, 1.23)$ \\
Three-year & $-3.44\; (-12.16, 9.78)$ & $1.09\; (-6.10, 8.69)$ & $1.90\; (-3.21, 6.35)$ & $-1.57\; (-4.43, 2.03)$ \\
\bottomrule\end{tabular}\end{table}

\FloatBarrier

\section{Sensitivity to the matching requirement}
\label{sec:s-matching}
The main matched comparison requires the same year, reviewer count and panel mean.
It retains 1,012 pairs, of which 698 have both citation outcomes available, and yields a
difference of $+0.17$ percentile points ($-2.60$ to $2.76$).

As a sensitivity analysis, we allow panel means to differ by at most 0.25 raw score
points, keeping exact matches on year and reviewer count. This maximum allowed difference
is called a caliper. Within each year and reviewer-count group, we first match submissions with the fewest eligible matches. Each takes the remaining eligible submission with the closest panel mean, without replacement; ties are resolved by submission identifier. The relaxed
requirement retains 1,773 pairs, of which 950 have both outcomes available, and yields
$+1.93$ points ($-0.44$ to $4.29$). The estimate remains uncertain and describes a different
matched sample. It also leaves a difference in panel means: the standardized mean
difference is 0.179 across all pairs and 0.159 among pairs with both outcomes available.
The exact-match comparison has identical years, reviewer counts and panel means
within each pair.

Standardized mean differences use the pre-match pooled standard deviation. Intervals resample the fixed pairs; they do not include uncertainty in
which pairs were formed.

\FloatBarrier

\section{Citation-window checks}
\label{sec:s-window} Because most submissions here circulate as preprints,
part of a submission's citation window can precede the panel's decision. Across the
two-year rejection windows, 5.5\% of citations carry a database date before the
official decision. The share is highest for minority support (8.0\%) and lowest for
the highest-score rule (4.7\%), indicating that this exposure differs across rules. Bibliographic
dates carry no precision indicator and a publication date is not the date a citation first
appeared, so we did not build a decision-anchored window on them. Counting every citation
in a boundary year first inside and then outside the window moves the variance rule's citation gain
from $1.91$ to $1.98$ at a budget of 10\%. These are two scenarios rather than bounds,
because percentiles re-rank and both arms move together; the spread is small relative to the interval width, so the calendar window is retained.

\FloatBarrier

\section{Year-weight diagnostic}
\label{sec:s-weights}
The observed-outcome contrast pools each rule's available selections. Consequently, year weights depend on which selected outcomes are observed. As a point-estimate diagnostic, we instead calculate the selected-minus-lottery difference within each edition and average those differences using the same budget weights. At a 10\% budget, the minority-support difference changes from $-1.12$ to $-1.47$ points in the two-year window and from $-1.06$ to $-1.51$ in the three-year window. This diagnostic has no newly estimated interval and does not establish equivalence under common year weights. It does not address selection bias within an edition.

\FloatBarrier

\section{Implementation details}
\label{sec:s-implementation}
Ranking scores are rounded to twelve decimal places; ties are ordered by submission identifier.

\FloatBarrier

\section{Equivalence-margin sensitivity}
\label{sec:s-margin}
To put the margin on a standardized scale, we use the sample standard deviation (SD) of citation percentiles among eligible rejections with observed outcomes, pooled across editions within each analysis window. Each submission's percentile uses the original within-edition reference distribution. The reference SD is 26.28 for the two-year window (7,411 submissions) and 26.30 for the three-year window (4,220 submissions). Five percentile points correspond to 0.19 SD in either window; 0.2 SD corresponds to 5.26 points. The standardizer is descriptive and held fixed, rather than an estimate of the standard error of a paired rule contrast. Conventional effect-size labels provide scale context, not a validated threshold for policy importance \citep{williams2014,lakens2018}.

We require the 95\% interval
half-width to fall below five points before evaluating equivalence at a given budget.
On the main window the half-width is
8.48 and 5.65 points at budgets of 1 and 2\%, and 3.49 and 2.30 at
5 and 10\%. At the two smallest budgets the interval is wider than the chosen margin, so equivalence is not evaluated at these budgets; positive differences from the lottery
can still be estimated at these budgets. The criterion describes the precision of each comparison.

We reuse the same 2,000 bootstrap draws and report every budget, including those below the original precision gate. Table~\ref{tab:s-margin} shows whether the 90\% interval lies strictly inside each symmetric margin. The minimum margin is the larger absolute endpoint of that interval; the margin must exceed it. This is a description of the existing interval, not a new chosen threshold. The complete grid from 0.5 to 10 points in steps of 0.25 is included with the aggregate
results \studyrepository.
\begin{table}[ht]\centering\small
\caption{Sensitivity of minority-support equivalence to the margin. All values except budget and SD labels are percentile points. Yes means that the 90\% bootstrap interval lies within the specified bounds; this is the interval criterion described in Methods. All comparisons concern observed outcomes.}
\label{tab:s-margin}
\resizebox{\textwidth}{!}{\input{supplement/tableS4.tex}}
\end{table}

At a 10\% budget, both windows meet the interval criterion at five points and at 0.2 SD, but neither meets it at three points. At a 5\% budget, neither window meets any of these three criteria. The five-point margin is held fixed across the reported comparisons. The standardized-effect calculation provides scale context and was not used to choose the margin.

\FloatBarrier

\section{Complete allocation-rule comparisons}
\label{sec:s-contrasts}
Table~\ref{tab:s-yields} reports citation gains against the lottery at each budget. Table~\ref{tab:contrast} compares the rules directly with minority-support ranking. Table~\ref{tab:s-budget-contrasts} reports the paired panel-mean and highest-score comparisons with minority-support ranking at every budget.

\begin{table}[t]
\centering
\caption{Citation gains under each rule, two-year window. Mean citation percentile of the selected set
minus that of a lottery over the same pool, in percentile points; positive is better than
chance. Intervals are 95\% percentile bootstrap over submissions, 2,000 resamples.
At budgets of 1 and 2\% the minority-support interval half-width exceeds the $\pm 5$ margin, so equivalence is not evaluated at these budgets, although their intervals can identify positive gains (Supplementary Section~\ref{sec:s-margin}). The
bar-clearing comparison matches submissions with the same year, reviewer count and panel mean; it has no budget. Its estimate uses pairs with citation outcomes available for both submissions.}
\label{tab:s-yields}
\small
\setlength{\tabcolsep}{3pt}
\begin{tabular}{lcccc}
\toprule
Budget & 1\% & 2\% & 5\% & 10\% \\
\midrule
Panel mean & $10.69$ & $8.27$ & $6.99$ & $6.63$ \\
 & ($2.73$, $15.98$) & ($3.88$, $14.14$) & ($3.68$, $10.00$) & ($4.11$, $8.66$) \\
Highest score & $9.23$ & $8.20$ & $6.11$ & $5.61$ \\
 & ($2.07$, $16.29$) & ($2.56$, $13.52$) & ($2.81$, $9.38$) & ($3.59$, $8.47$) \\
Score variance & $0.71$ & $1.74$ & $0.58$ & $1.91$ \\
 & ($-6.28$, $8.90$) & ($-3.79$, $6.92$) & ($-2.40$, $4.54$) & ($-0.26$, $4.29$) \\
Minority support & $3.77$ & $1.48$ & $-2.51$ & $-1.12$ \\
 & ($-6.62$, $10.35$) & ($-4.44$, $6.86$) & ($-5.91$, $1.06$) & ($-3.58$, $1.02$) \\
\midrule
\multicolumn{5}{l}{\emph{Bar-clearing indicator, matched, two-year window}} \\
\quad Exact match & \multicolumn{4}{l}{$0.17$ \; ($-2.60$, $2.76$)} \\
\multicolumn{5}{l}{1,012 pairs; 698 with both outcomes available} \\
\bottomrule
\end{tabular}
\end{table}

\begin{table}[ht]
\centering
\caption{Each ranking rule compared with minority-support ranking, budget of 10\%. Positive means the rule
selects work later cited above what minority support selects. Each difference uses the same
bootstrap draws for both terms, so the interval reflects the correlation between rules
evaluated on one pool. The four samples are nested inside one another, not independent. All
comparisons are exploratory, and each interval is read on its own without any correction for
testing several at once.}
\label{tab:contrast}
\small
\resizebox{\textwidth}{!}{%
\begin{tabular}{lcccc}
\toprule
 & Two-year & Two-year & Two-year & Three-year \\
Rule minus minority support & 2017--2024 & through 2023 & through 2022 & through 2022 \\
\midrule
Panel mean & $7.75$ & $8.77$ & $11.00$ & $12.15$ \\
 & ($4.35$, $11.09$) & ($5.10$, $12.74$) & ($6.72$, $15.73$) & ($7.86$, $16.87$) \\
Highest score & $6.72$ & $7.32$ & $8.89$ & $10.08$ \\
 & ($4.34$, $10.24$) & ($4.68$, $11.58$) & ($5.98$, $14.34$) & ($7.23$, $15.50$) \\
Score variance & $3.02$ & $4.47$ & $4.21$ & $4.64$ \\
 & ($0.51$, $5.98$) & ($1.76$, $7.94$) & ($1.18$, $8.37$) & ($1.46$, $8.88$) \\
Lottery & $1.12$ & $0.65$ & $0.22$ & $1.06$ \\
 & ($-1.02$, $3.58$) & ($-1.88$, $3.64$) & ($-2.74$, $3.85$) & ($-1.91$, $4.81$) \\
\bottomrule
\end{tabular}}
\end{table}

At a budget of 5\%, panel-mean ranking exceeds minority-support ranking by $9.51$
percentile points ($4.39$ to $13.98$). At 10\%, highest-score ranking after adjustment
for panel mean and reviewer count differs from minority-support ranking by $-0.17$
points ($-1.78$ to $2.39$). These comparisons use the main two-year sample. At 10\%, score-variance ranking is $4.72$ points below panel-mean ranking ($1.18$ to $7.59$); its gain over the lottery through 2023 is $3.82$ points ($1.15$ to $6.77$). At 5\%, its difference from minority-support ranking is $3.09$ points ($-0.26$ to $7.28$).

\begin{table}[ht]\centering\small
\caption{Panel-mean and highest-score ranking compared with minority-support ranking at all four budgets in the main two-year sample. Entries are paired differences in citation percentile points (95\% bootstrap intervals). Positive values favor the listed rule.}
\label{tab:s-budget-contrasts}
\begin{tabular}{lcc}\toprule
Budget & Panel mean minus minority support & Highest score minus minority support \\
\midrule
1\% & $6.92$ ($-3.50$, $18.03$) & $5.46$ ($-3.01$, $17.62$) \\
2\% & $6.79$ ($0.27$, $15.50$) & $6.72$ ($0.06$, $14.06$) \\
5\% & $9.51$ ($4.39$, $13.98$) & $8.62$ ($4.01$, $12.92$) \\
10\% & $7.75$ ($4.35$, $11.09$) & $6.72$ ($4.34$, $10.24$) \\
\bottomrule\end{tabular}
\end{table}

\FloatBarrier

\section{Including the 2020 edition}
\label{sec:s-2020}
We repeat the ranking analyses after adding the original 2,593 submissions from 2020,
using the same rule definitions, citation data, within-year percentiles and 2,000 paired
bootstrap draws. The edition adds 1,526 eligible rejections, of which 1,482 (97.1\%)
have citation outcomes. The two-year sample then includes 12,151 eligible rejections,
with 8,893 observed outcomes (73.2\%); the three-year sample includes 6,438,
with 5,702 observed outcomes (88.6\%). At a 10\% budget, the number selected increases
from 686 to 755 in the two-year window and from 301 to 370 in the three-year window.

Panel-mean ranking retains a higher citation gain than minority-support ranking in
both windows (Table~\ref{tab:s-2020}). Minority support remains close to the lottery.
The two-year score-variance interval now lies just above zero, while its advantage over
minority support remains positive. At the smallest two-year budget (1\%), the paired
panel-mean versus minority-support interval includes zero. The complete four-budget
results and missing-outcome scenarios are retained in the accompanying analysis files.

The exact-matching analysis gains no pairs from 2020. Among its eligible rejections,
957 have at least one score above the estimated acceptance threshold and 569 do not.
Within reviewer-count groups, the two groups have no identical panel means. The original
1,012 pairs, including 698 with both outcomes observed, therefore remain unchanged.

\begin{table}[ht]
\centering\small
\caption{Ranking comparisons after including 2020, at a budget of 10\%. Values are citation-percentile differences with 95\% paired bootstrap intervals among observed outcomes. The two-year window covers 2017--2024 and the three-year window covers 2017--2022, both including 2020.}
\label{tab:s-2020}
\begin{tabular}{lcc}\toprule
Comparison & Two-year & Three-year \\
\midrule
Panel mean vs lottery & $7.39$ ($5.09$, $9.36$) & $11.57$ ($8.56$, $14.23$) \\
Highest score vs lottery & $5.25$ ($3.62$, $8.02$) & $7.59$ ($5.55$, $11.66$) \\
Score variance vs lottery & $2.32$ ($0.21$, $4.74$) & $3.85$ ($1.27$, $6.94$) \\
Minority support vs lottery & $-0.81$ ($-3.22$, $1.08$) & $-0.73$ ($-3.69$, $2.00$) \\
Adjusted highest score vs lottery & $-0.84$ ($-2.97$, $1.53$) & $-0.91$ ($-3.41$, $2.11$) \\
Panel mean vs minority support & $8.20$ ($5.00$, $11.35$) & $12.29$ ($8.45$, $16.38$) \\
Score variance vs minority support & $3.13$ ($1.06$, $6.00$) & $4.58$ ($1.98$, $8.13$) \\
\bottomrule
\end{tabular}
\end{table}

\FloatBarrier

\section{Sensitivity to sample and missing-outcome assumptions}
\label{sec:s-missing}
\begin{table}[ht]
\centering
\caption{Sensitivity analyses, budget of 10\%. Panel A repeats the comparison on the three-year
window, which restricts the sample to editions through 2022. Panel B refits every rule on
the pool that also contains withdrawn submissions, as a sensitivity analysis.
Panel C varies the assumption about submissions with no citation record, recomputing the
lottery benchmark under each; these are assumptions and not bounds. Panel D reports
worst-case bounds for minority support: the first assumes nothing about unresolved
submissions beyond the observed percentile scale, the second assumes they would have landed
at or below the first quartile of their year, and the third is an outer envelope that
re-ranks against the full cohort and is not sharp.}
\label{tab:robust}
\small
\resizebox{\textwidth}{!}{%
\begin{tabular}{lcccc}
\toprule
\multicolumn{5}{l}{\emph{A. Three-year window}} \\
\midrule
Panel mean & \multicolumn{4}{l}{$11.08$ \; ($7.90$, $14.10$)} \\
Highest score & \multicolumn{4}{l}{$9.02$ \; ($6.44$, $13.29$)} \\
Score variance & \multicolumn{4}{l}{$3.58$ \; ($0.48$, $7.07$)} \\
Minority support & \multicolumn{4}{l}{$-1.06$ \; ($-4.81$, $1.91$)} \\
\midrule
\multicolumn{5}{l}{\emph{B. Pool including withdrawn submissions, two-year window}} \\
\midrule
Panel mean & \multicolumn{4}{l}{$5.51$ \; ($3.25$, $7.84$)} \\
Highest score & \multicolumn{4}{l}{$5.99$ \; ($3.54$, $8.49$)} \\
Score variance & \multicolumn{4}{l}{$2.27$ \; ($-0.30$, $4.55$)} \\
Minority support & \multicolumn{4}{l}{$0.98$ \; ($-1.55$, $3.14$)} \\
\midrule
\multicolumn{5}{l}{\emph{C. Missingness assumptions, two-year window}} \\
Unresolved assigned & lowest percentile & pool mean & zero citations & \\
\midrule
Panel mean                  & $9.88$  & $5.05$  & $7.88$  & \\
Highest score               & $6.40$  & $3.98$  & $5.37$  & \\
Score variance              & $2.77$  & $1.31$  & $2.19$  & \\
Minority support            & $-1.52$ & $-0.69$ & $-1.16$ & \\
\midrule
\multicolumn{5}{l}{\emph{D. Worst-case bounds for minority support, two-year window}} \\
\midrule
No assumption on unresolved & \multicolumn{4}{l}{$[-31.09,\; 29.87]$} \\
Unresolved at or below the year's first quartile & \multicolumn{4}{l}{$[-6.93,\; 4.21]$} \\
Outer envelope, full-cohort re-rank & \multicolumn{4}{l}{$[-44.73,\; 44.13]$} \\
\bottomrule
\end{tabular}}
\end{table}

Table~\ref{tab:robust} reports the alternative citation window, the expanded pool including withdrawals,
and the missing-outcome assumptions. With no restriction on unresolved outcomes beyond
the observed percentile scale, minority support has bounds of $[-31.09,29.87]$.
Allowing the reference percentiles to change with the missing outcomes gives the wider
outer envelope of $[-44.73,44.13]$. Assuming unresolved submissions fall at or below
their year's first quartile narrows the bounds to $[-6.93,4.21]$, still extending beyond
the five-point equivalence margin.

A separate calculation starts by assigning each unresolved submission its year's
resolved-rejection mean, giving minority support a gain of $-0.69$ points. At a budget
of 10\%, 231 of its 686 selections are unresolved. Raising the assigned outcomes of
these selections by $2.20$ points brings the estimate to zero; a rise of $18.13$ points
brings it to $+5$, and a fall of $13.73$ points brings it to $-5$. Thus, changing the
sign requires a smaller shift than reaching a five-point advantage. This calculation
changes only the rule's unresolved selections under the stated scenario; it is not a
guarantee against other patterns of missing outcomes.

\FloatBarrier

%% file: supplement/tableS1.tex
\begin{tabular}{lrrr}\toprule
Characteristic & All eligible & Outcome available & Outcome unavailable \\
\midrule
\multicolumn{4}{l}{\textit{A. Outcome coverage: percentages of eligible rejections in each row}} \\
All editions & 10,625 & 7,411 (69.8\%) & 3,214 (30.2\%) \\
2017 & 244 & 239 (98.0\%) & 5 (2.0\%) \\
2018 & 491 & 472 (96.1\%) & 19 (3.9\%) \\
2019 & 917 & 892 (97.3\%) & 25 (2.7\%) \\
2021 & 1,735 & 1,689 (97.3\%) & 46 (2.7\%) \\
2022 & 1,525 & 928 (60.9\%) & 597 (39.1\%) \\
2023 & 2,226 & 1,224 (55.0\%) & 1,002 (45.0\%) \\
2024 & 3,487 & 1,967 (56.4\%) & 1,520 (43.6\%) \\
\midrule
\multicolumn{4}{l}{\textit{B. Characteristics within each column}} \\
Reviewers, mean (SD) & 3.73 (0.63) & 3.68 (0.62) & 3.85 (0.63) \\
Bar-clearing indicator, n (\%) & 6,582 (61.9\%) & 4,582 (61.8\%) & 2,000 (62.2\%) \\
\bottomrule\end{tabular}

%% file: supplement/tableS2.tex
\begin{tabular}{lrrr}\toprule
Edition & Eligible rejections & Panel mean, mean & Panel mean, SD \\
\midrule
2017 & 244 & 4.74 & 0.97 \\
2018 & 491 & 4.65 & 0.90 \\
2019 & 917 & 4.80 & 0.87 \\
2021 & 1,735 & 5.01 & 0.78 \\
2022 & 1,525 & 4.72 & 0.97 \\
2023 & 2,226 & 4.67 & 0.96 \\
2024 & 3,487 & 4.76 & 0.95 \\
\bottomrule\end{tabular}

%% file: supplement/tableS4.tex
\begin{tabular}{lrrrrrr}\toprule
Window & Budget & 90\% interval & Minimum margin & $\pm3$ & $\pm5$ & $\pm0.2$ SD \\
\midrule
Two-year & 1\% & $(-5.21, 9.08)$ & 9.08 & No & No & No \\
Two-year & 2\% & $(-3.66, 5.95)$ & 5.95 & No & No & No \\
Two-year & 5\% & $(-5.47, 0.54)$ & 5.47 & No & No & No \\
Two-year & 10\% & $(-3.30, 0.63)$ & 3.30 & No & Yes & Yes \\
Three-year & 1\% & $(-8.04, 8.47)$ & 8.47 & No & No & No \\
Three-year & 2\% & $(-6.44, 6.00)$ & 6.44 & No & No & No \\
Three-year & 5\% & $(-7.60, 0.71)$ & 7.60 & No & No & No \\
Three-year & 10\% & $(-4.25, 1.36)$ & 4.25 & No & Yes & Yes \\
\bottomrule\end{tabular}